\documentclass{mem}
\usepackage{natbib}\usepackage{txfonts}\usepackage{balance}
\usepackage{graphicx}
\usepackage[a4paper]{hyperref}
\idline{75}{282}
\begin{document}
\def\teff{$T\rm_{eff }$}
\def\kms{$\mathrm {km s}^{-1}$}

\title{
Accretion onto black holes and relativistic jets
}

   \subtitle{}

\author{
T.\,Belloni\inst{1} 
          }

  \offprints{T. Belloni}

\institute{
Istituto Nazionale di Astrofisica --
Osservatorio Astronomico di Brera, Via E. Bianchi 46,
I-23807 Merate, Italy
\email{tomaso.belloni@brera.inaf.it}
}

\authorrunning{Belloni}

\titlerunning{Accretion and relativistic jets}

\abstract{
Relativistic jets from Active Galactic Nuclei are known since decades, but the study of the connection between accretion and ejection in these systems is hampered by the long time scales associated to these events. The past decade has seen a rapid advancement due to the observation of similar radio jets in galactic X-ray binaries, where the time scales are much shorter. A clear connection between accretion and ejection has been found for these systems, together with a solid characterization of the phenomenological properties of their outbursts. This wealth of new results has led to a detailed comparison between X-ray binaries and AGN, from which a number of correlations and scaling laws has been established. Here I briefly review the current observational status.

}
\maketitle{}

\section{Introduction}

Relativistic jets from Active Galactic Nuclei (AGN) are known since decades. It is clear that they are associated with accretion onto the central black hole. However, this relation is difficult to study as the time scales for accretion (and ejection) events are of the order of years or centuries. 
The situation has changed in the past years due to the discovery that X-ray binaries, in particular black-hole binaries, always show evidence of the presence of a jet save for when they are in their soft state (see Fender 2006 for a review). The shorter time scales for Galactic events, combined with the availability of all-sky monitor instruments and flexible high-energy missions such as the Rossi X-Ray Timing Explorer (RXTE), has led to an rapid increase in our knowledge of the accretion-ejection connection. 

\begin{figure*}[t!]
\begin{tabular}{cc}
\resizebox{7.2 true cm}{!}{\includegraphics[clip=true]{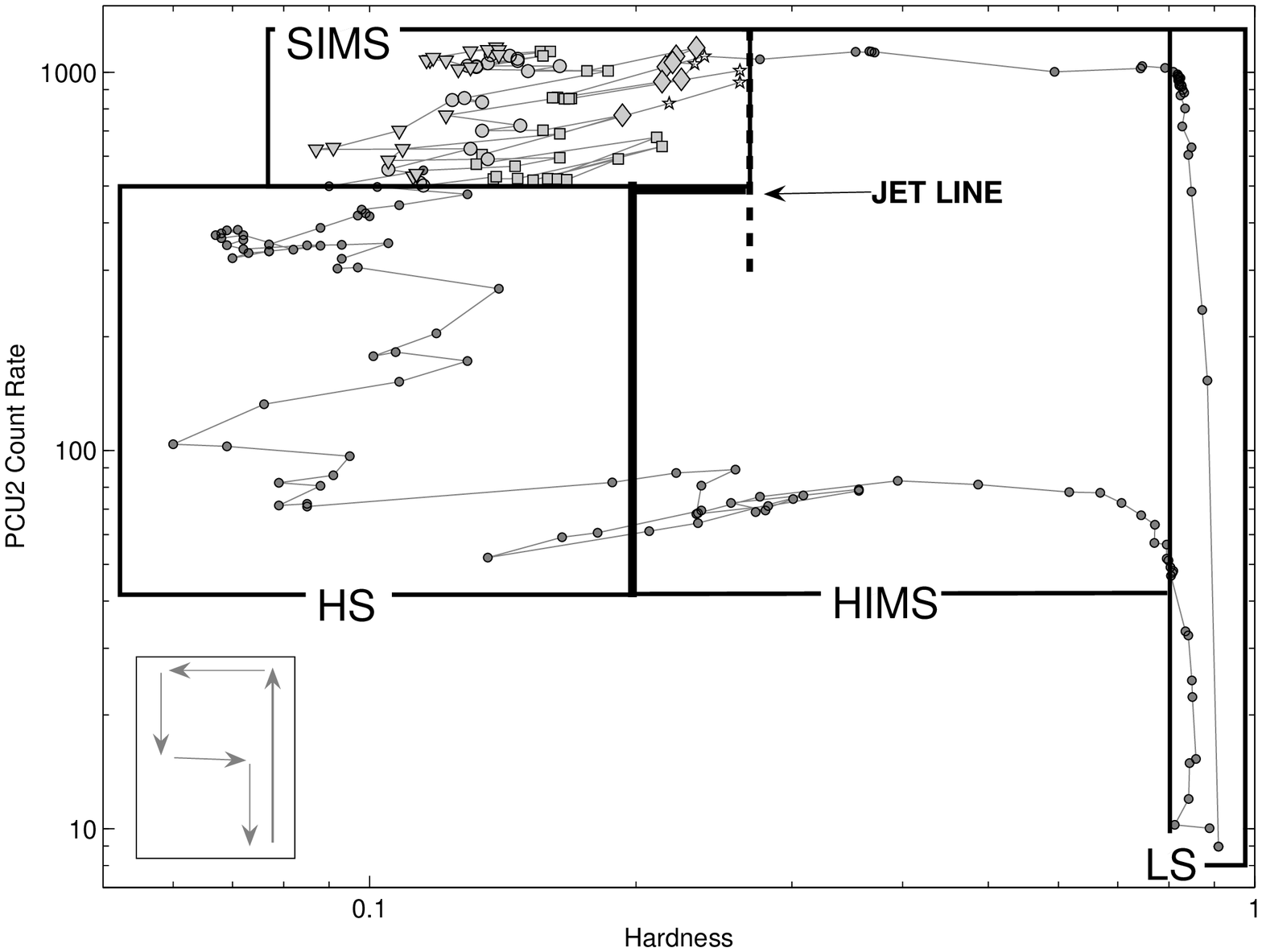}} &
\resizebox{5.0 true cm}{!}{\includegraphics[clip=true]{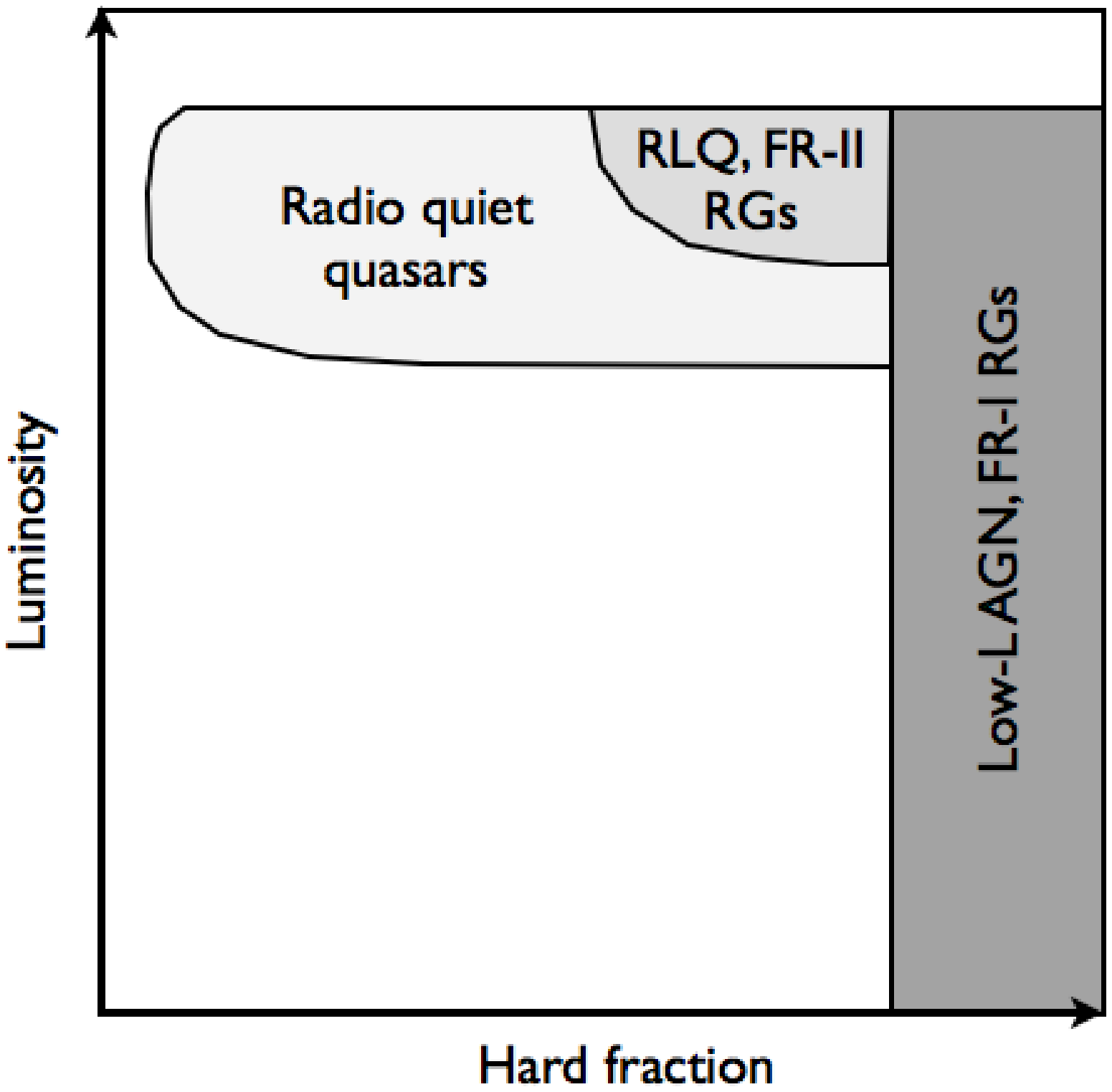}}  \\
\end{tabular}
\caption{\footnotesize
Left panel: hardness-intensity diagram of GX 339--4 (from Belloni 2006). Hard spectra correspond to points on the right. The path of evolution of an outburst is marked by arrows. The lines indicate the transitions between different spectral/timing states (labeled after Homan \& Belloni 2005, Belloni et al. 2006). The dashed line is the ``jet line''.
Right panel: corresponding diagram showing the likely position of different classes of AGN [adapted from K\"ording et al. 2006]. Here hardness is defined between optical and X-rays. Darker areas correspond to higher radio emission.
}
\label{xradio}
\end{figure*}

\section{Jets in X-ray binaries}

Two types of jets are observed in X-ray binaries.  The so-called ``steady'' or compact jets are associated to states with the hardest X-ray emission (see Homan \& Belloni 2005 and Remillard \& McClintock 2006 for two alternative state definition paradigms), show a self-absorbed radio spectrum and are estimated to have moderate velocities (Fender, Belloni \& Gallo 2004, Miller-Jones, Fender \& Nakar  2006). The radio flux shows a strong correlation with the X-ray flux, of the form $L_R \propto L_X^{0.7}$ (Gallo, Fender \& Pooley 2003).
These states correspond to the initial and final parts of the outbursts of transient systems and are mostly associated to accretion rates below about a tenth of the Eddington luminsity. On the other hand, when the sources are in their soft state, dominated by a cool thermal accretion disk, no radio detection has been obtained so far, with upper limits a factor of $\sim$50 lower than the hard-state detections. Interestingly, neutron-star binaries also show similar radio emission in their hard state, correlated with the X-ray flux with a steeper index, but also at fluxes systematically lower by a factor of $\sim$30 (Migliari \& Fender 2006).

The other type of jet corresponds to the fast transient ejections that led to the detection of superluminal radio jets (see Fender 2006 for a review). Here the estimated Lorentz factors are much higher (see Miller-Jones et al. 2006) and the radio spectrum is steep. From GRS 1915+105 the first source in which these jets were observed (Mirabel \& Rodr\'\i guez 1994), it was soon found that the ejections corresponded to precise events in the X-ray band (see Fender \& Belloni 2004). Through observations of more conventional black-hole transients, it emerged that this is the case for all systems: the ejection time was traced back to coincide with the transition between hard and soft states (Fender, Belloni \& Gallo 2004). In the left panel of Fig. \ref{xradio} the evolution of the prototypical transient GX 339--4 during an outburst is sketched in the hardness-intensity plane (harder spectra correspond to points on the right). As the outburst proceeds, the diagram is followed counter-clockwise and the source passes through the different states (Homan \& Belloni 2005; Belloni et al. 2005). The line whose crossing corresponds to the jet ejection is the dotted line at the top, also called ``jet line'' (Fender, Belloni \& Gallo 2004). This diagram contains X-ray intensity and hardness, but it also corresponds to a clear subdivision between radio-loud (LS and HIMS) and radio-quiet points (SIMS, HS), indicated by the thick line in Fig \ref{xradio} (left panel). The position of the lines which mark the transitions from one state to the other are based on the properties of fast time variability (see Homan \& Belloni 2005; Belloni 2006).

The general picture that emerges is the following (see Fender, Belloni \& Gallo 2004). As a transient leaves its quiescence and starts becoming brighter, it remains in its hard state and the compact jet becomes brighter following the X-ray/radio correlation, but its velocity is moderate ($\Gamma\sim 1-2$). As the spectrum starts becoming softer, at the top of the right branch in Fig. \ref{xradio}, a transition which can last as little as one day and whose cause is not understood, a thermal accretion disk starts appearing in the X-ray spectrum. The Lorentz factor of the jet rises fast as the ``jet line'' is approached, leading to a fast jet impinging on the slower one. This causes the bright ejection, after which the jet production is suppressed. All that is visible is the `bullet'' fired at the crossing of the line, but no nuclear radio emission. This corresponds to the soft (and soft-intermediate, SIMS) state. On the reverse transition (soft to hard) at much lower luminosities, all that happens is that the compact radio jet is formed again, but no fast ejection takes place (see Kalemci et al. 2006). The systems goes back to its hard/quiescent state. If small back-transitions take place at higher luminosities, repeated ejections can be observed.

\section{Connections to AGN}

\begin{figure*}[t!]
\begin{tabular}{cc}
\resizebox{6 true cm}{!}{\includegraphics[clip=true]{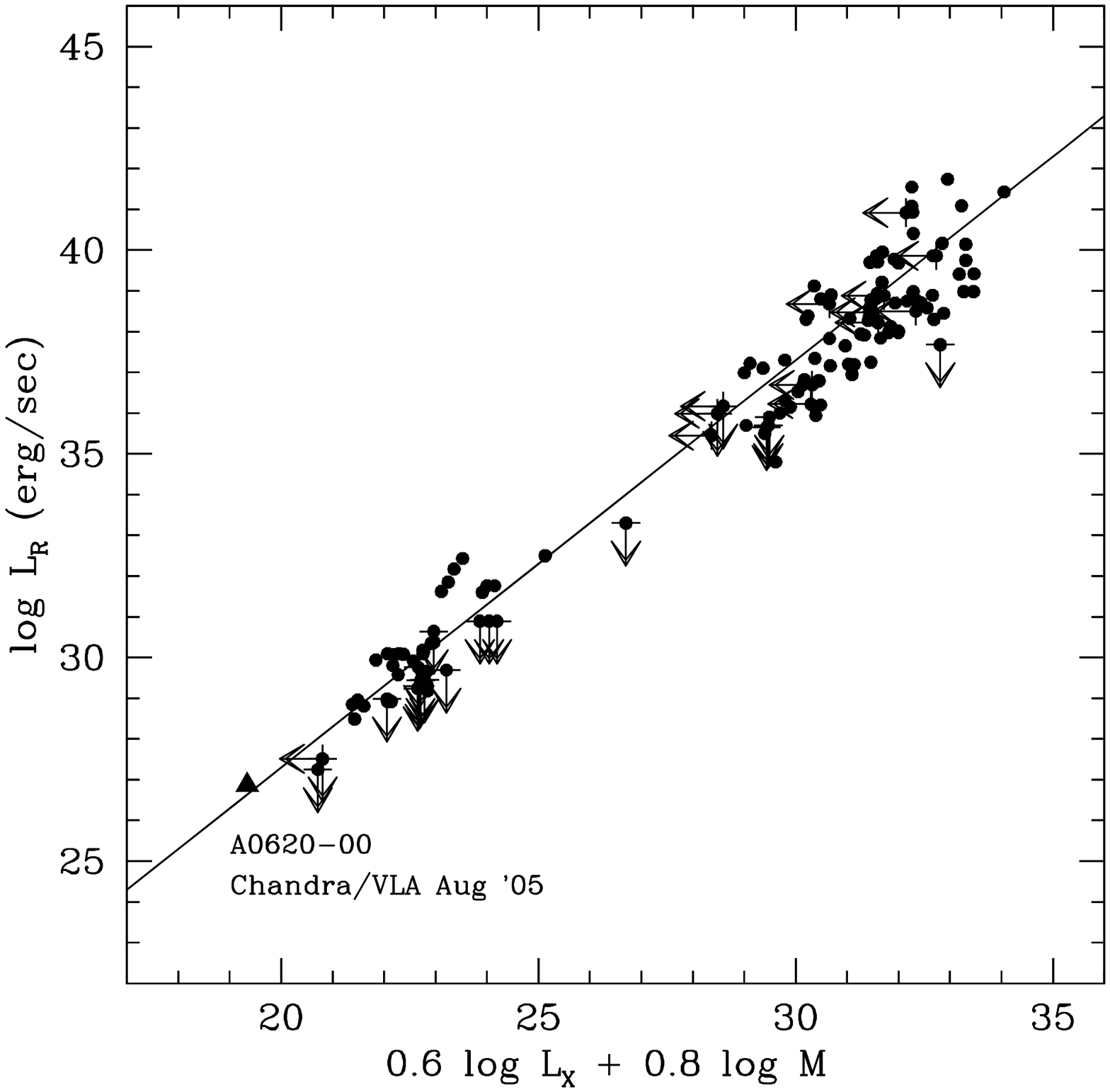}} &
\resizebox{7.1 true cm}{!}{\includegraphics[clip=true]{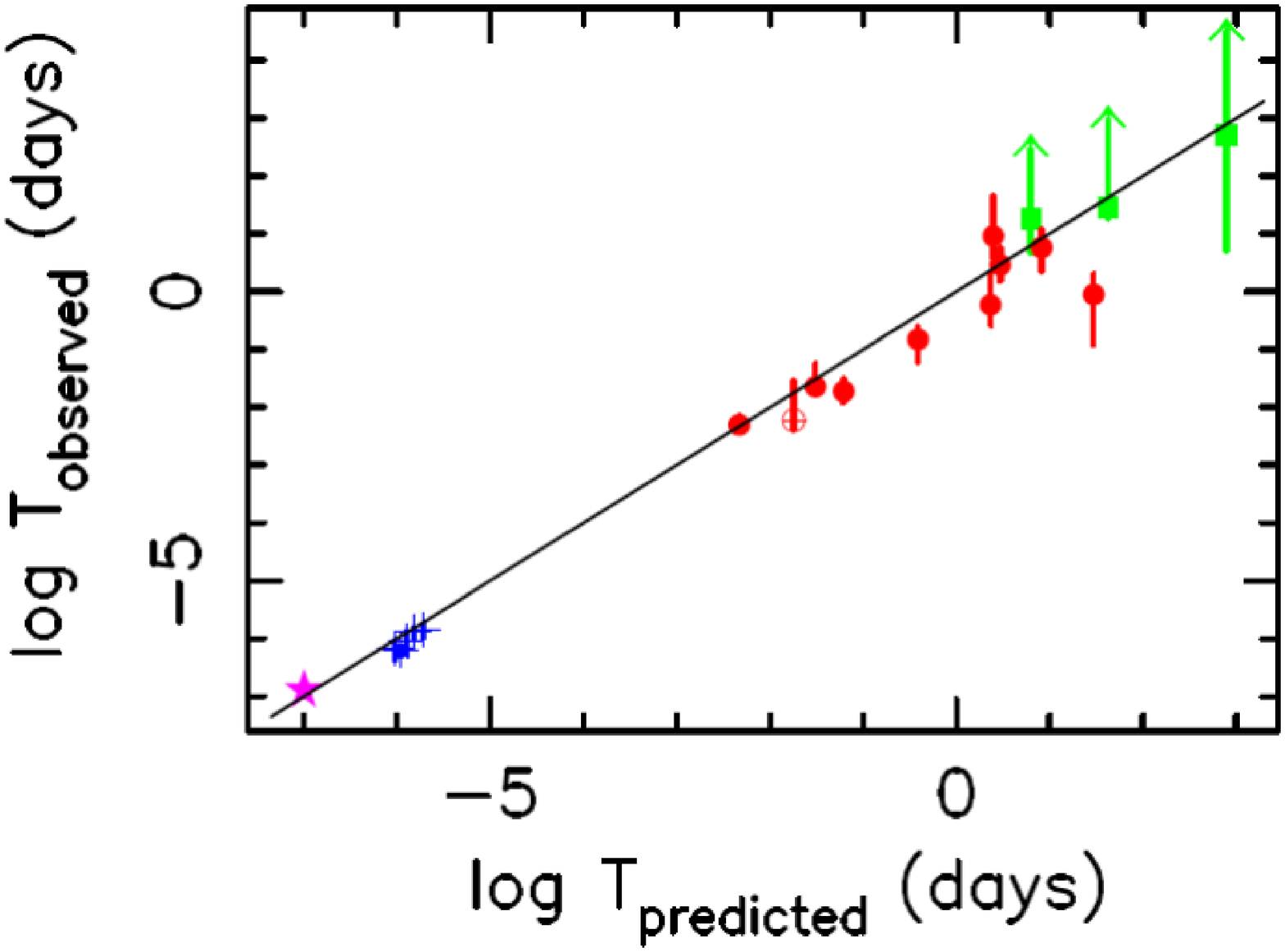}}  \\
\end{tabular}
\caption{\footnotesize
Left panel: correlation between radio luminosity and X-ray luminosity (with a mass correction factor) which includes AGN, black-hole binaries in their hard state and one black-hole binary in quiescence (original from Merloni, Heinz \& Di Matteo 2003, this plot from Gallo et al. 2006). The line is the best fit (see Gallo et al. 2006).
Right panel:
correlation between the observed time scale associated to a break in the power density spectrum of a sample of AGN and black-hole binaries observed with RXTE and the value predicted by a quadratic relation in black-hole mass and a linear relation in bolometric luminosity (see McHardy et al. 2006).
}
\label{correlations}
\end{figure*}

Once a more solid knowledge of the properties of stellar-size systems is available, it is possible to extend it to galaxy-sized sources. As mentioned, a clear correlation between radio and X-ray flux was found for X-ray binaries (Gallo, Fender \& Pooley 2003, Migliari \& Fender 2006). This correlation was extended to a sample of AGN by Merloni, Heinz \& Di Matteo (2003). Since to compare sources on largely different scale, one needs to apply a correction for the different masses (which are more or less the same for stellar-sized objects), this correlation was dubbed ``fundamental plane". Recently, Gallo et al. (2006) have extended it to low flux levels with the radio detection of a black-hole binary in quiescence (A 0620--00). The fundamental plane can be seen in the left panel of Fig. \ref{correlations}: it now covers 15 orders of magnitude! Of course this plot includes only sources in the hard state, since in the soft state no radio detection has been obtained. 

The diagram in the left panel of Fig. \ref{xradio} is extremely important for the classification and study of black-hole binaries. In the case of AGN, the long time scales imply that on human-lifetime scales most of the sources will occupy a single position in such a diagram. Of course, with a large sample of sources, a similar diagram should be recovered. This would help to understand and identify accretion states in AGN and their relation to radio loudness. K\"ording et al. (2006) constructed a similar diagram using as measure of hardness the disk fraction, i.e. the fraction of flux that can be attributed to a thermal accretion disk. From this, they constructed a sketch that shows the possible association between AGN classes and black-hole binary states (see right panel of Fig. \ref{xradio}). Currently, the situation concerning AGN is not completely clear, but we now have a template for comparison.

As mentioned above, a very important aspect of the emission of galactic X-ray binaries is their fast time variability, which is paramount to identify different states and in particular state transitions (some transitions involve very small shifts in Fig. \ref{xradio}, but major changes in the fast variability).
More than eleven years of RXTE observations of AGN have allowed the accumulation of a large and unique database for the study of aperiodic variability of AGN. Although the level of details is nowhere near that available for X-ray binaries, from the associated power spectra, characteristic time scales can be identified. These can be compared with those typically detected from X-ray binaries, which of course are much faster and usually below 1--10 seconds. A scaling with mass of the central object is expected, but it is also known from X-ray binaries that these time scales vary as a function of luminosity and hence of accretion rate. Indeed, a simple time-mass relation was not found in the past. 
Recently, McHardy et al. (2006) have shown that the observed characteristic time scales of large and small systems in their soft states (associated to breaks in the continuum components in the power spectra, see also Belloni, Psaltis \& van der Klis 2002), can be related, indicating that the accretion process is the same on black holes of all scales. The observed characteristic time was found to scale as $T_B\propto M^2 L^{-1}$, where $M$ is the mass of the black hole and $L$ is the luminosity. This means that the time associated to the observed break in the power spectra is $T_B\propto M/\dot m_E$, where $\dot m_E$ is the accretion rate in units of Eddington accretion rate. 
This result shows that the physical properties of the accretion onto black-hole binaries and of AGN are fundamentally the same and opens new possibilities for the theoretical interpretation of such time scales, which are not at all understood. The result of McHardy et al. (2006) for soft-state objects is being complemented by the addition of hard-state objects (K\"ording et al. 2007), which appear to follow a similar relation with different normalization. 

\section{Conclusions}

The past decade has seen a remarkable advancement in our knowledge of X-ray binaries. It is now clear that the jet is a fundamental element in these systems, ignoring which would impede serious theoretical modeling. These results are now bouncing back to the field of accreting supermassive black holes in AGN. The comparison of the slow-evolving and relatively faint (but numerous) and the fast and bright (but few) X-ray binaries is a very promising avenue to reach an understanding of their common processes.
 
\begin{acknowledgements}
The author acknowledges financial contribution from contract ASI-INAF  I/023/05/0
\end{acknowledgements}

\bibliographystyle{aa}

\end{document}